\documentclass[conference]{IEEEtran}
\ifCLASSINFOpdf
  \usepackage[pdftex]{graphicx}
\else
  \usepackage[dvips]{graphicx}
\fi
\usepackage{amsmath}
\ifCLASSOPTIONcompsoc
  \usepackage[caption=false,font=normalsize,labelfont=sf,textfont=sf]{subfig}
\else
  \usepackage[caption=false,font=footnotesize]{subfig}
\fi

\usepackage{stfloats}
\usepackage{url}
\usepackage{color}
\usepackage{gensymb}
\usepackage{comment}
\usepackage{graphicx}
\usepackage{amsmath}
\usepackage{gensymb}
\usepackage{multirow}
\usepackage{tabularx}
\usepackage{adjustbox}

\usepackage{algorithm}
\usepackage{algorithmic}
\usepackage{xcolor}
\usepackage{comment}
\usepackage{soul}

\begin{document}
%
\title{A Unified Hardware Architecture for Convolutions and Deconvolutions in CNN}

\author{
\IEEEauthorblockN{Lin Bai, Yecheng Lyu and Xinming Huang}
\IEEEauthorblockA{Worcester Polytechnic Institute\\
\{lbai2, ylyu, xhuang\}@wpi.edu}
}


\maketitle

\begin{abstract}
Deconvolution plays an important role in the state-of-the-art convolutional neural networks (CNNs) for the tasks like semantic segmentation, image super resolution, etc. 
In this paper, a scalable neural network hardware architecture for image segmentation is proposed. By sharing the same computing resources, both convolution and deconvolution operations are handled by the same process element array. In addition, access to on-chip and off-chip memories is optimized to alleviate the burden introduced by partial sum. As an example, SegNet-Basic has been implemented using the proposed unified architecture by targeting on Xilinx ZC706 FPGA, which achieves the performance of 151.5 GOPS and 94.3 GOPS for convolution and deconvolution respectively. This unified convolution/deconvolution design is applicable to other CNNs with deconvolution.  
\end{abstract}

%
\IEEEpeerreviewmaketitle

\section{Introduction}
In recent years, CNNs have been widely used in computer vision applications such as classification \cite{ref:vgg_2014}, object detection \cite{ref:fasterrcnn_2015}
, and semantic segmentation \cite{ref:fcn_2015}. However, a CNN usually requires intensive computations, which limits its applicability on embedded devices. To address this issue, FPGA based accelerators \cite{ref:eyeriss_2016} 
have been proposed so that CNNs can be important on real-time embedded systems. As a key operation in many neural networks, deconvolution has been widely used in the state-of-the-art CNNs especially for semantic segmentation\cite{ref:fcn_2015}, image super resolution\cite{ref:super_res_2017}, and image denoising\cite{ref:denoise_2017}. Through a learnable way, deconvolution extrapolates new information from input feature maps, which outperforms other interpolation algorithms such as nearest neighbor and bi-cubic interpolation. However, unlike hardware acceleration for convolution, much less attention has been paid on deconvolution. Due to the fact that 
deconvolution may become the bottleneck in speed if only convolution has been accelerated, there is a urgent need to optimize the deconvolution operation on FPGAs.

\section{Related Work}

Lately tremendous research progress has been made on high-performance and low-power CNN accelerators. In \cite{ref:utku_2017}, the authors proposed a novel architecture for process element array, which dramatically reduced the external memory bandwidth requirements by intensive data reuse and outperforms the systolic-like structures\cite{ref:chen_2016}. A high-throughput CNN accelerator design was implemented in \cite{ref:xuechao_2017}, where a comprehensive design space exploration on top of accurate models was deployed to determine the optimal design configuration. 

Comparing to the accelerator design for convolution, that of deconvolution has not been thoroughly investigated. 
Liu et al. proposed an CNN architecture where convolution and deconvolution were accelerated on the system separately\cite{ref:shuanglong_2018}. This architecture is not efficient enough because in most CNNs, convolution and deconvolution do not work in parallel. A high performance deconvolution module in \cite{ref:xinyu_2017} used reverse looping and stride hole skipping techniques, but with the penalty of additional hardware resources and latency. An unified systolic accelerator was developed in \cite{ref:fcnelement_2018}, which divided the deconvolution into two steps. Firstly, it multiplied one vector with the kernel and then stored the temporary matrices in on-chip memory. Next, it added the overlaps of temporary matrices. This method increased the on-chip BRAM access and introduced unnecessary data storage. Consequently both power consumption and computation latency grew.

To address the issues mentioned above, we analyze the deconvolution properties and fit it into our proposed process element array so that both convolution and deconvolution can be handled by sharing the same on-chip resources. The contributions of our work are summarized as follows:

\begin{itemize}
	\item A novel process element structure is proposed so that both convolution and deconvolution are supported without extra circuit.
	\item A scalable CNN accelerator architecture is proposed, in which data rate and processing speed are configurable depending on the on-chip resources and bandwidth of the target FPGA. 
	\item SegNet-Basic has been implemented on Xilinx ZC706 FPGA efficiently. Its throughput are 151.5 GOPS and 94.3 GOPS for convolution and deconvolution respectively, which outperforms most of the existing FPGA accelerators.
\end{itemize}

The rest of paper is organized as follows. Section \ref{sec:Backgrounds} introduces the background concept. The hybrid-optimization strategies are described in Section \ref{sec:Optimization}. Hardware architecture and its implementation results of SegNet-Basic are discussed in Section \ref{sec:Architecture}-\ref{sec:Results}. In the end, Section \ref{sec:Conclusions} concludes the paper. 

\section{Backgrounds}\label{sec:Backgrounds}
\subsection{CNN architecture for segmentation}
\begin{figure}[htbp]
	\centering
	\includegraphics[width=0.6\columnwidth]{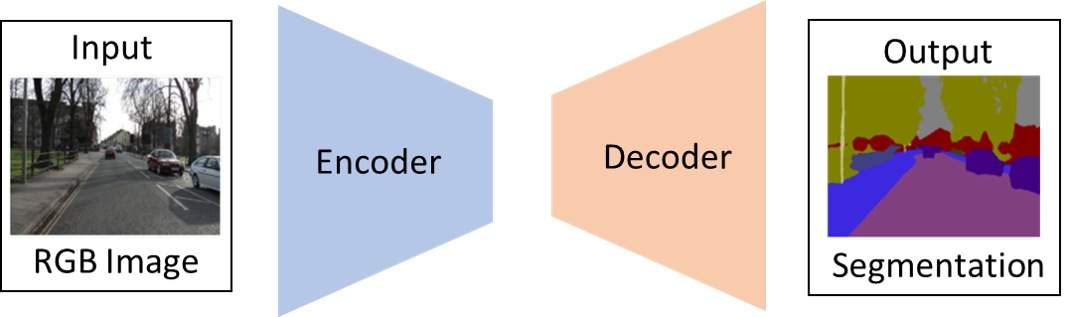}
	\caption{Typical CNN structure for semantic segmentation}
	\label{fig:seg_arch}
\end{figure}

Typical segmentation neural network architecture consists of an encoder and a decoder as shown in Fig.~\ref{fig:seg_arch}. The encoder is used to extract the features from the input image and the decoder generates the segmented output. In CNNs like SegNet and U-Net, decoder reverses their encoder, while others replace decoder with smaller one. Additional connections between encoder and decoder were introduced to SegNet \cite{ref:segnet_2017} and U-Net \cite{ref:u-net_2015}. Without the extra connection, CNN still works but with a small decrease of precision. Usually, the encoder is comprised of convolutional layers and the decoder consists of deconvolutional layers.

\subsection{Deconvolution}
\begin{figure}[htbp]
	\centering
	\includegraphics[width=0.7\columnwidth]{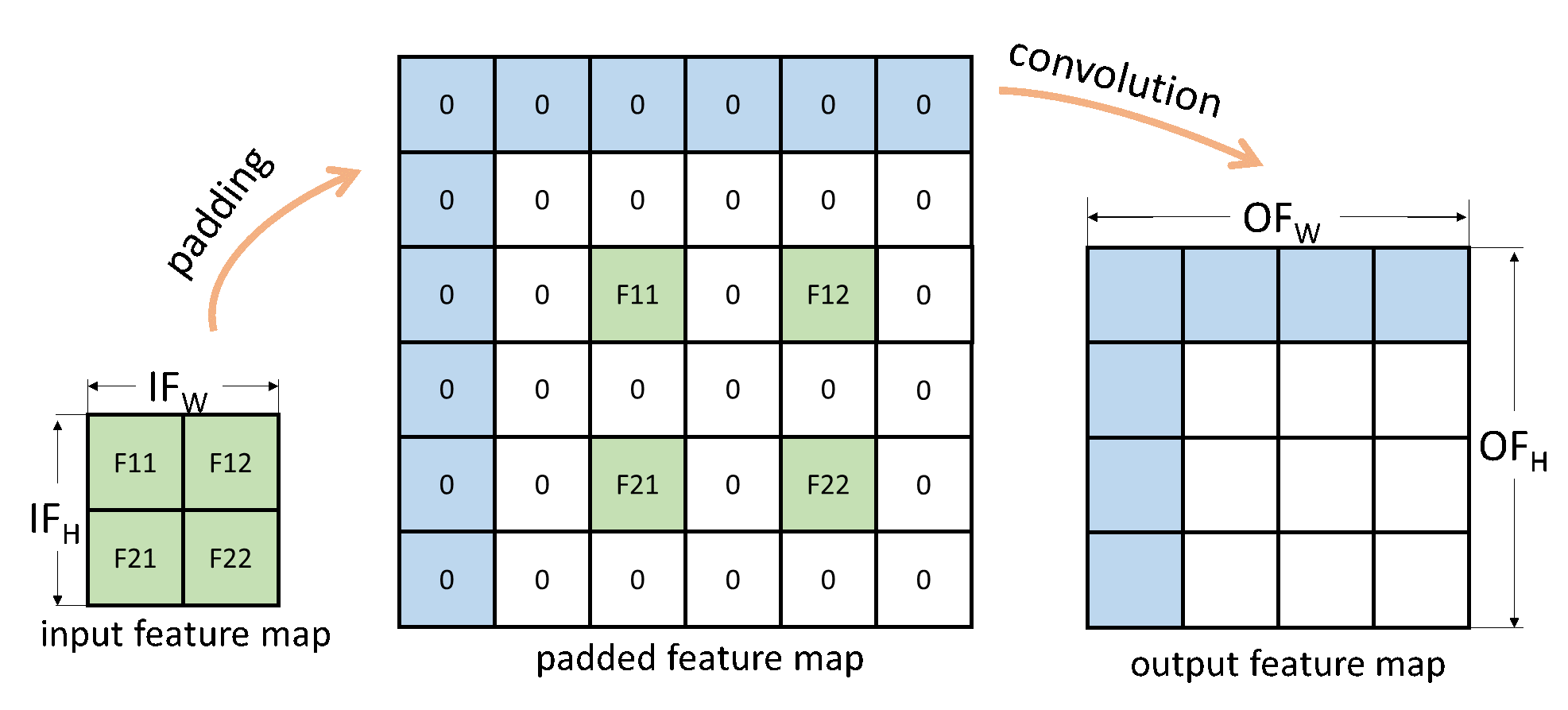}
	\caption{Deconvolution operation by padding and convolution}
	\label{fig:deconv_op}
\end{figure}

Deconvolution, also called transposed convolution, is a learnable method to perform upsampling. If a convolution unit is directly reused for deconvolution, it consists of the following two steps: 1) padding the input feature map and 2) applying convolution on the padded feature map, as indicated in Fig.~\ref{fig:deconv_op}. After padding, an input feature map with size $I\!F_W\times I\!F_H$ is expanded into $(2\cdot I\!F_W+1)\times (2\cdot I\!F_H+1)$, and consequently the output feature map size becomes $(2\cdot I\!F_W-1)\times (2\cdot I\!F_H-1)$. In order to get exactly twice the size, extra padding for upper row and left column (highlighted in blue in Fig.~\ref{fig:deconv_op}) is needed. 

\section{Optimization}\label{sec:Optimization}
Because of the limited resources on an FPGA, a high performance CNN accelerator has to be deeply optimized on memory access and data transfer while maximizing the resource utilization.
\subsection{Loop optimization}
To efficiently map the convolution loops, three loop optimization techniques, loop unrolling, loop tiling and loop interchange, have been considered to customize the computation and communication patterns of the accelerator. Loop unrolling is the parallelism strategy for certain convolution loops, which demands more multipliers. Loop tiling determines the partition of feature maps, and consequently determines the required size of on-chip memory. Loop interchange decides the computation order of the convolution loops \cite{ref:optmloop_2017}.

\begin{figure}[htbp]
	\centering
	\includegraphics[width=0.8\columnwidth]{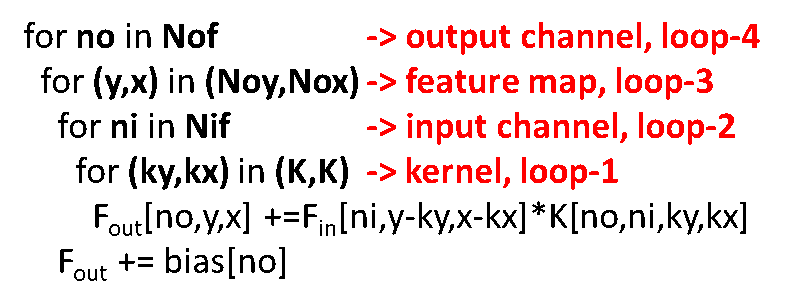}
	\caption{Four levels of convolution loops}
	\label{fig:conv_loop}
\end{figure}

After carefully judging all three optimization methods, our optimization strategy is to unroll loop-1, and partially unroll loop-2, loop-4, and apply loop tiling on depth of input feature maps (Fig.~\ref{fig:conv_loop}). In order to jointly optimize with deconvolution, loop-1 is fully unrolled (more details about this will be explained in the next section). In order to reduce the number of partial sums and data transfer, loop-2 must be unrolled as much as possible. However, as the large amount of multipliers are required, loop-2 is only partially unrolled. Further consideration has been taken for on-chip memory access minimization. Therefore, loop-4 is unrolled because of pixel reuse. As the partial sum in loop-2 are stored in BRAM, no more overhead to the off-chip memory is added.

\subsection{Deconvolution}\label{sec:opt_deconv}
Fig.~\ref{fig:deconv_op} presents 
the naive way to implement the deconvolution on hardware. As it can be seen, too many operations are wasted in multiplication by zeros.

\begin{figure}[htbp]
	\centering
	\includegraphics[width=0.7\columnwidth]{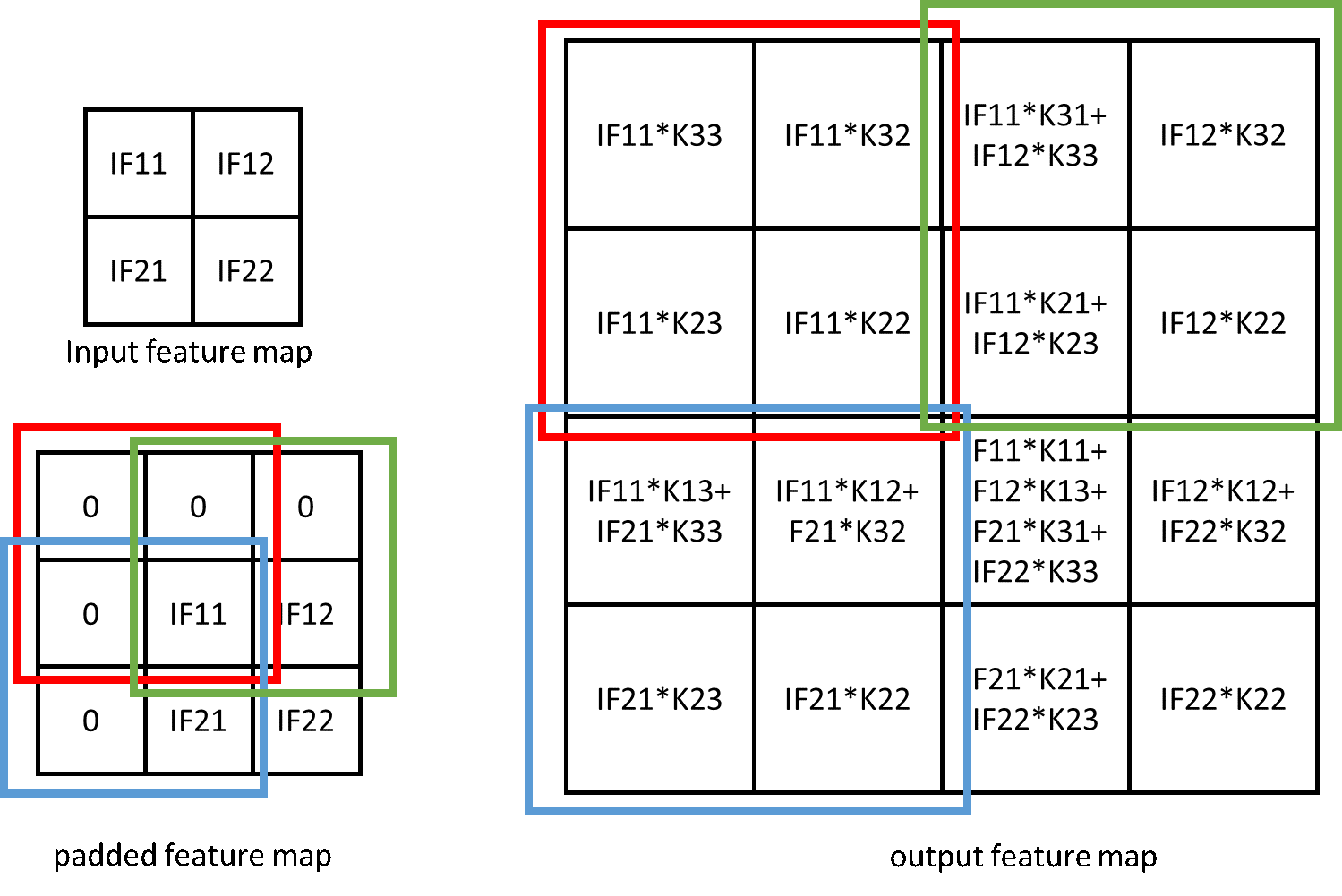}
	\caption{Optimization of deconvolution}
	\label{fig:deconv_result}
\end{figure}

 The mathematical expression of the $2\times 2$ feature map deconvolution is given in Fig.~\ref{fig:deconv_result}. Based on equation (\ref{eq:deconv_1}-\ref{eq:deconv_4}) for the deconvolution, we conclude most of the redundant multiplications can be avoided. The procedure is summarized into three steps: 1) padding the input feature map if size doubling is expected; 2) scanning the padded input feature map by a $2\times 2$ sliding window; 3) applying deconvolution for each patch using kernel as in equation (\ref{eq:deconv_1}-\ref{eq:deconv_4}). Three examples are highlighted by colored squares in padded feature map and output feature map in Fig.~\ref{fig:deconv_result}.

\begin{align}
    O\!F_{11}\!=&I\!F_{11}\!\cdot\!K_{11}\!+\!I\!F_{12}\!\cdot\!K_{13}\!+\!I\!F_{21}\!\cdot\!K_{31}\!+\!I\!F_{22}\!\cdot\!K_{33}\label{eq:deconv_1}\\
    O\!F_{12}\!=&I\!F_{12}\!\cdot\!K_{12}\!+\!I\!F_{22}\!\cdot\!K_{32}\\
    O\!F_{21}\!=&I\!F_{21}\!\cdot\!K_{21}\!+\!I\!F_{22}\!\cdot\!K_{23}\\
    O\!F_{22}\!=&I\!F_{22}\!\cdot\!K_{22}\label{eq:deconv_4}
\end{align}

According to TensorFlow, during deconvolution, the kernel should be rotated by 180\degree. To make the figure easier to understand, we assume that the kernel has been rotated already.

\subsection{Quantization method}
A fine-tuning with quantization constraint method \cite{ref:yecheng_tcas_2019} is employed in our design. It effectively diminishes the negative impact of brute-force quantization while introducing more non-linearity. Different from the ordinary quantization method \cite{ref:google_quant}, we quantize the weights and bias before storage. This quantization method does not require modification of the TensorFlow source code.

\section{Hardware Architecture}\label{sec:Architecture}
\begin{figure}[htbp]
	\centering
	\includegraphics[width=0.9\columnwidth]{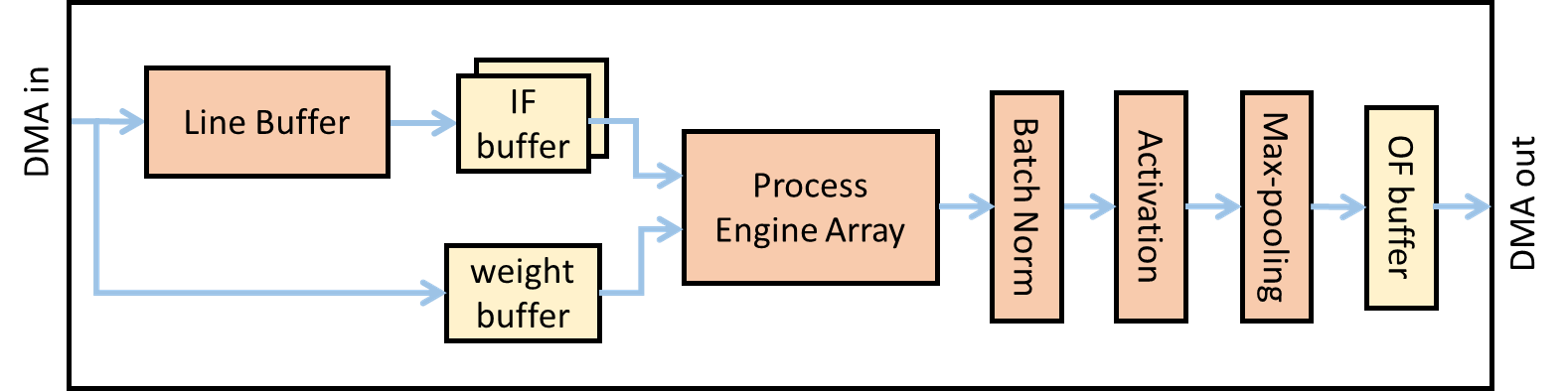}
	\caption{Overview of hardware architecture}
	\label{fig:acce_arch}
\end{figure}

The overview of hardware architecture is shown in Fig.~\ref{fig:acce_arch}. Line buffer converts the convolution into matrix multiplication by re-organizing the input image. The process element array multiplies input image by the weights. After batch normalization, activation and pooling, the output feature map is stored in Output Featuremap (OF) buffer.

\subsection{Line buffer}

\begin{figure}[htbp]
	\centering
	\includegraphics[width=0.9\columnwidth]{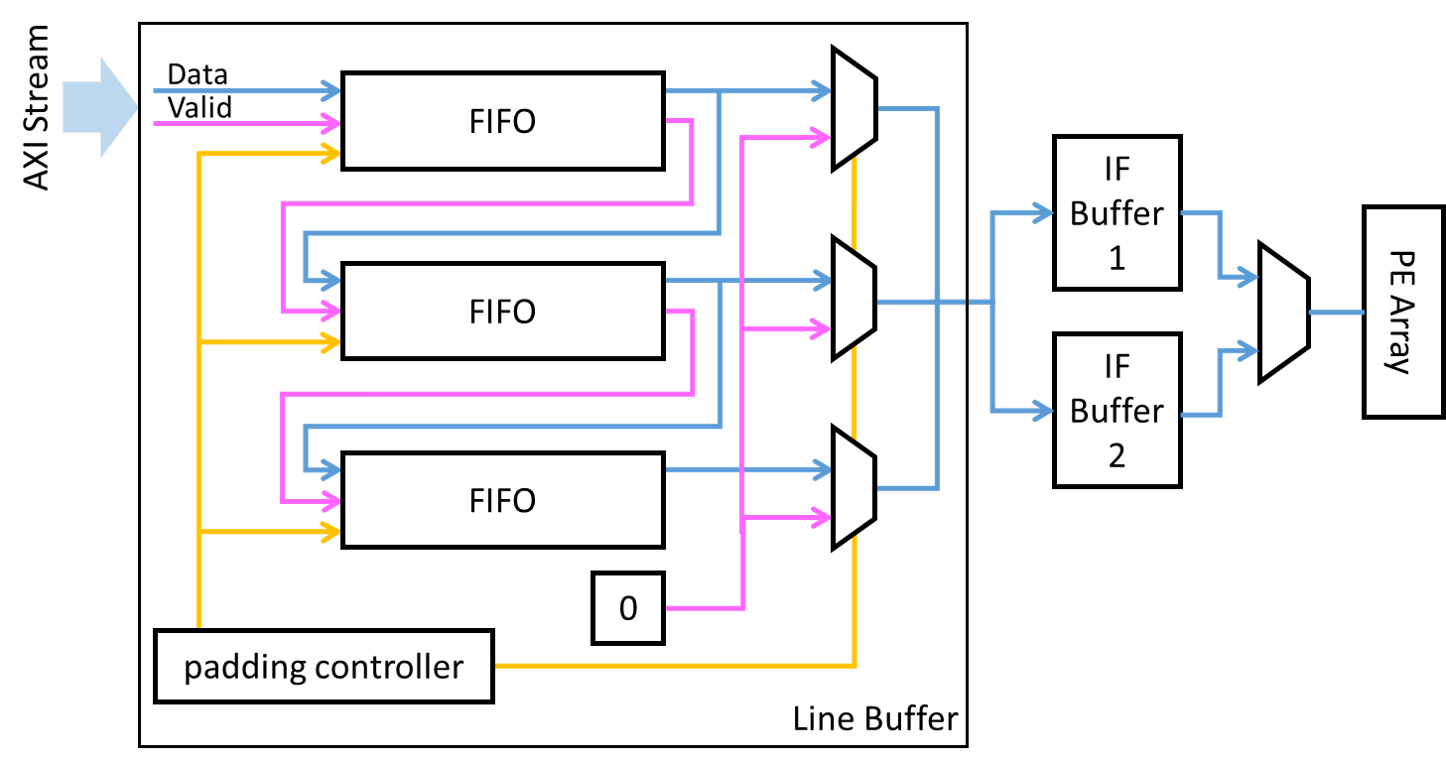}
	\caption{Block diagram of line buffer}
	\label{fig:zero_pad}
\end{figure}

A line buffer is designed to build the expected sliding window and to perform zero-padding for convolution. In the proposed accelerator, line buffer bridges the AXI DMA and Input Featuremap (IF) buffer (Fig.~\ref{fig:zero_pad}). Data and valid signals from AXI Stream interface are inserted into cascaded FIFOs. Extra logic is added to generate the status signals (including empty and full signals) of each FIFO, so that the line buffer is able to fit feature maps with different sizes. Padding controller decides when to push the data into each FIFO and when to output zero padding according to the pre-loaded padding mode.

We choose not to buffer one entire feature map in on-chip memory, because the buffer size would be restricted by the limited on-chip memory and this could result inefficient buffer usage when feature map size drops. Therefore, only part of the input feature map is loaded into IF buffer and consequently this requires different zero-padding modes. Hence we provide different pre-loaded work modes for this line buffer.

\subsection{Process element array}
\begin{figure}[htbp]
	\centering
	\includegraphics[width=0.6\columnwidth]{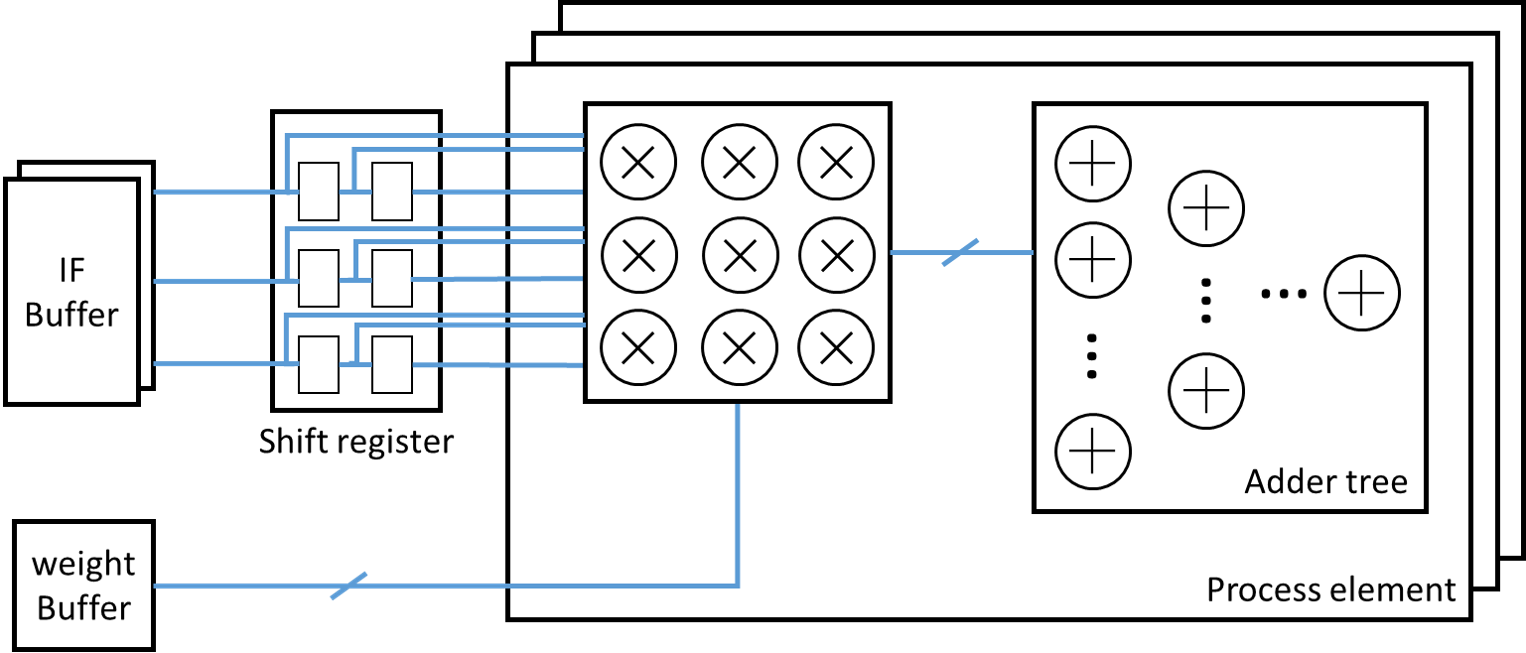}
	\caption{PE array structure and its connection to buffers}
	\label{fig:conv}
\end{figure}

The input feature map is stored in the IF buffers in form of $3\times 1$ vector, which reduces the BRAM consumption. To rebuild the $3\times 3$ / $2\times 2$ sliding window for convolution/deconvolution, a shift register is placed between IF buffer and process element arrays. Each process element array consists of multiple process elements as in Fig.~\ref{fig:conv}, usually in a power of 2. This number is scalable, depending on the bandwidth of platform. Each process element comprises of a 9-multiplier array and an adder tree to sum up the products.

\subsubsection{Convolution}
During convolution, $3\times 3$ sliding windows and their corresponding weights are transmitted into process element array. In the process element, they are multiplied and summed up.

\subsubsection{Deconvolution}
As discussed in Section \ref{sec:opt_deconv}, the deconvolution for each $2\times 2$  patch consumes $9$ multiplications and $5$ additions. This fits well to the proposed process element. During deconvolution, the process element structure is reused but with a different data routing mechanism.

Different from convolution who outputs one pixel after another, in the deconvolution mode, one process element generates $4$ pixels in parallel. Considering the bitwidth of OF buffer, these output pixels are fed into buffer in serial.

\subsection{Pooling}
The pooling operation in CNNs is either max pooling or average pooling for $2\times 2$ sliding windows. Another line buffer and shift register are utilized to generate $2\times 2$ sliding windows. Its work mode can be determined prior to compilation or configured on-the-fly.

\subsection{Batch normalization and activation}
During inference, the batch normalization is downgraded into a multiplication with an addition. We simply absorb it into the process element. Concerning to the activation function, our design supports both ReLU and LeakyReLU.

\subsection{System controller}
The system controller determines: 1) the data flow such as when to trigger process element array, which data to be assigned into buffers, when to start data transfer to DDR, and 2) the setting for each component mentioned above, like the padding mode of line buffer module, work mode of process element arrays, and whether to bypass activation or pooling modules. It is implemented as a FSM. These settings are pre-defined and pre-loaded into register files so that FSM only has to read and execute sequentially.

\begin{table*}[!htb]
    \centering
    \caption{Resource comparison for CNN implementations on FPGA}
    \label{tab:comparison}
    \begin{adjustbox}{width=0.9\textwidth}
    \begin{tabular}{|c|c|c|c|c|c|c|} 
        \hline
        & \cite{ref:ref_13} & \cite{ref:ref_18} &  \cite{ref:shuanglong_2018} & \cite{ref:xinyu_2017} & Ours\\
        \hline
        FPGA & ZC7Z045 & ZC7Z045 & ZC7Z045 & ZC7Z020 & ZC7Z045 \\
        \hline
        Clock (MHz) & 150& 100 & 200 & 100 & 220 \\
        \hline
        Precision & 16-bit fixed & 16-bit fixed & 16-bit fixed & 12-bit fixed & 8-bit fixed\\
        \hline
        Network & VGG16-SVD  & VGG19  & U-Net  &  GAN & SegNet-basic \\
        \hline
        Operation & CONV & CONV & CONV+DECONV & DECONV & CONV+DECONV \\
        \hline
        Performance & \multirow{2}{*}{187} & \multirow{2}{*}{229} & 125 (CONV) & \multirow{2}{*}{2.6} & 151.5 (CONV) \\
        (GOPS) &  &  & 29 (DECONV)  &  & 94.3 (DECONV)\\
        \hline
        Resource Efficiency & \multirow{2}{*}{0.207(CONV)} &  \multirow{2}{*}{0.254} & 0.14 (CONV) & \multirow{2}{*}{0.012} & 0.168 (CONV) \\
        (GOPS/DSP)  & & & 0.033 (DECONV) & & 0.104 (DECONV)\\
        \hline
    \end{tabular}
    \end{adjustbox}
\end{table*}

\section{Implementation Considerations of SegNet-Basic}\label{sec:SegNet}
The encoder part of SegNet-Basic includes $4$ convolution layers and  $3$ max pooling layers. The decoder part has $2$ convolution layers and $3$ deconvolution layers. The total parameter size for inference is about 42Mb, with 8-bit quantization for feature maps and weights. The accelerator is designed using Simulink and the HDL Coder toolbox. Our target platform Xilinx ZC706 contains 900 DSP slices and 19.2Mb Block RAMs.

\section{Results and Discussion}\label{sec:Results}

\begin{figure}[htbp]
	\centering
	\includegraphics[width=0.55\columnwidth]{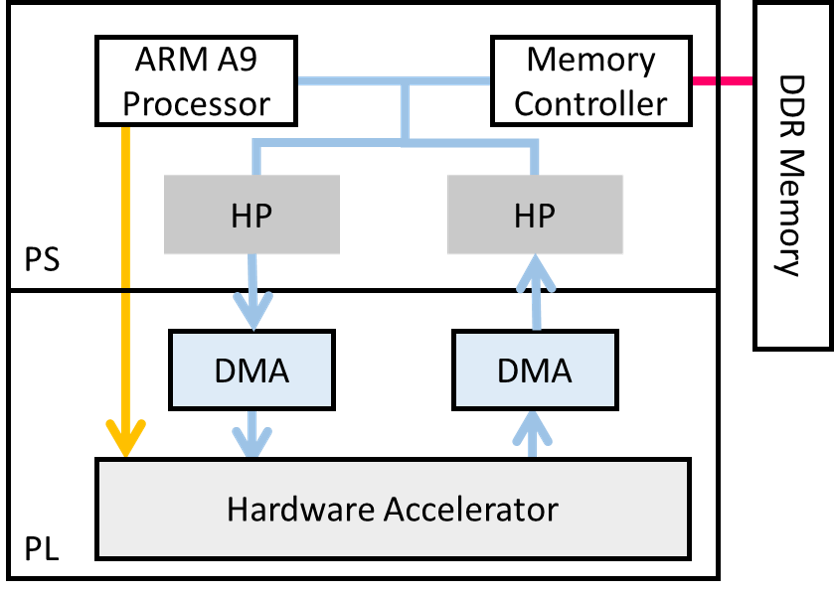}
	\caption{The hardware architecture of CNN accelerator}
	\label{fig:sys_arch}
\end{figure}

The test setup of SegNet-basic hardware accelerator is demonstrated in Fig.~\ref{fig:sys_arch}. In the Zynq platform, hardware accelerator is loaded as a peripheral of ARM A9 processor. Two Direct Memory Access (DMAs) move the data between accelerator and DDR memory. The input images and parameters are pre-loaded into memory and transferred to PL by DMAs. This CNN accelerator clock frequency is 220MHz. Its total resource consumption is summarized in Tab.~\ref{tab:resource}.

\begin{table}[htbp]
    \centering
    \caption{Resource consumption}
    \label{tab:resource}
    \begin{tabular}{|c|c|c|c|}
        \hline
        LUTs & Registers & BRAMs & DSPs \\
        \hline
        16579 (8\%) & 25390 (6\%) & 537 (99\%) & 576 (64\%) \\
        \hline
    \end{tabular}
\end{table}
Comparing to other implementations (in Tab.~\ref{tab:comparison}), our design achieves better performance in case of deconvolution. Due to the sharing architecture, a better balance on both performance and resource efficiency for convolution and deconvolution is obtained. However, in order to support both operations, the architecture is not deeply optimized for convolution specifically. Therefore, the convolution performance is not as high as that from deeply optimized implementation in \cite{ref:ref_18}.

\subsection{Scalability}
Scalability is represented by the number of process element arrays in the accelerator. It is balance of bandwidth and computation capability. In SegNet-Basic, the number of process element arrays is set to 1. This means the input and output data bitwidth is 64. If higher bandwidth is supported, higher performance is possible.

\subsection{Latency of operations}
In order to compare the latency, we perform convolution and max pooling on a $90\times 120$ feature map (resulting a $45\times 60$ feature map) followed by deconvolution. We find the time for convolution and deconvolution are the same. The padding time difference is about $0.6\mu s$ due to different sizes of input feature maps. Considering pooling and ReLU, another $1.2\mu s$ is needed. Double buffering eliminates the data transfer time difference. Therefore, deconvolution saves about 3.2\% processing time if comparing to convolution plus max-pooling and ReLU.

\section{Conclusions}\label{sec:Conclusions}
In this paper, a scalable and configurable CNN accelerator architecture has been proposed by combining both convolution and deconvolution into single process element. The deconvolution operation is completed in one step and buffering of intermediate results is not needed. In addition, SegNet-Basic has been successfully implemented on Xilinx Zynq ZC706 FPGA that achieves the performance of 151.5 GOPS for convolution and 94.3 GOPS for deconvolution, which outperforms state-of-the-art segmentation CNN implementations.

\section*{Acknowledgment}
This work was supported by the Mathworks Inc.

\end{document}